# The MammoGrid Virtual Organisation - Federating Distributed Mammograms

Florida Estrella[a], Richard McClatchey[a], Dmitry Rogulin[a,b]

[a] University of the West of England (UWE), Bristol, United Kingdom

[b] European Centre for Nuclear Research (CERN), Geneva, Switzerland

**Abstract**

> The MammoGrid project aims to deliver a prototype which enables the effective collaboration between radiologists using grid, service-orientation and database solutions. The grid technologies and service-based database management solution provide the platform for integrating diverse and distributed resources, creating what is called a 'virtual organisation'. The MammoGrid Virtual Organisation facilitates the sharing and coordinated access to mammography data, medical imaging software and computing resources of participating hospitals. Hospitals manage their local database of mammograms, but in addition, radiologists who are part of this organisation can share mammograms, reports, results and image analysis software. The MammoGrid Virtual Organisation is a federation of autonomous multi-centres sites which transcends national boundaries. This paper outlines the service-based approach in the creation and management of the federated distributed mammography database and discusses the role of virtual organisations in distributed image analysis.

*Keywords:*

Medical Informatics; Grid; Service-Orientation

## 1. Introduction

The medical community has been exploring collaborative approaches for managing mammographic image data and the Grid technology [1] is a promising approach in enabling distributed analysis across medical institutions without the necessity for the clinicians to co-locate. The EU-funded MammoGrid project [2] aims to use existing Grids technologies in developing a European wide database of mammograms to support effective co-working among healthcare professionals across the EU. The project has been active since late 2002 and involves hospitals in the UK and Italy, medical imaging experts and academics with experience of implementing grid-based database solutions.

One of the deliverables of this project is a software prototype using open-source Grid middleware and service-based database management system that is capable of managing federated mammogram databases distributed across Europe. The proposed solution is a medical information infrastructure delivered on a service-based, grid-aware framework, encompassing geographical regions of varying protocols, lifestyles and diagnostic procedures.

The prototype will allow, among other things,
a) mammogram data mining,
b) diverse and complex epidemiological studies,
c) statistical and computer aided detection (CADe) analyses, and
d) deployment of versions of the image standardization software Standard Mammogram Form (SMF).

The MammoGrid collaboration is composed of the following partners:
- Mirada Solutions and the University of Oxford – medical image analysis expertise, acquisition system and the SMF software
- University of the West of England and the European Centre for Particle Physics (CERN) – Grid and database service provision
- Udine Hospital – including a database of several thousand mammograms
- Addenbrookes Hospital – sourcing a database of around 2,000 mammogram cases
- Universities of Pisa and Sassari – delivering Computer Aided Detection (CADe) software.

**2. The Grid Technologies**

The grid is defined as "*flexible, secure, coordinated resource sharing among dynamic collections of individuals, institutions and resources*" [1]. Geographically separated but working together to solve a problem, groups of people can be organised in collaborations, referred to as virtual organisations (VO), and use the shared resources of the grid.

The Grid can provide a virtual platform for large-scale, resource-intensive and distributed applications. It offers a connectivity environment allowing management and coordination of diverse and dispersed resources. The Grid enables access to increased storage and computing capacity providing mechanisms for sharing and transferring large amounts of data as well as aggregating distributed resources for running computationally expensive procedures. Another important point is that the Grid utilizes a common infrastructure based on open standards thus providing a platform for interoperability and interfacing between different Grid-based applications from the particular domain.

Grid technology can potentially provide medical applications an architecture for easy and transparent access to distributed heterogeneous resources (like data storage, networks, computational resources) across different organizations and administrative domains. The Grid offers a configurable environment allowing grid structures to be reorganized dynamically without disturbing any overall active Grid processing.

In particular the Grid can address some of the following issues relevant to medical domains:

*Data distribution:* The Grid provides a connectivity environment for medical data distributed over different sites. It solves the location transparency issue by providing mechanisms which permit seamless access to and the management of distributed data. These mechanisms include services which deal with virtualization of distributed data regardless of their location.

*Heterogeneity:* The Grid addresses the issue of heterogeneity by developing common interfaces for access and integration of diverse data sources. Such generic interfaces for consistent access to existing, autonomously managed databases that are independent of underlying data models are defined by the Global Grid Forum Database Access and Integration Services (GGF-DAIS) [3] working group. These interfaces can be used to represent an abstract view of data sources which can permit homogeneous access to heterogeneous medical data sets.

*Data processing and analysis:* The Grid offers a platform for transparent resource management in medical analysis. This allows the virtualization and sharing of all resources (e.g. computing resources, data storage, etc.) connected to the grid. For handling computationally intensive procedures (e.g. CADe), the platform provides automatic resource allocation and scheduling and algorithm execution, depending on the availability, capacity and location of resources.

*Security and confidentiality*: Enabling secure data exchange between hospitals distributed across networks is one of the major concerns of medical applications. Grid addresses security issues by providing a common infrastructure for secure access and communication between grid-connected sites. This infrastructure includes authentication and authorization mechanisms, among other things, supporting security across organizational boundaries.

*Standardization and compliance:* Grid technologies are increasingly being based on a common set of open standards (such as XML, SOAP, WSDL, HTTP etc.) and this is promising for future medical image analysis standards.

The next section discusses the MammoGrid Virtual Organization.

## 3. The MammoGrid Virtual Organisation

The MammoGrid Virtual Organisation (MGVO) is composed of three mammography centres – Addenbrookes Hospital (UK), Udine Hospital (Italy) and Oxford University (UK). These centres are autonomous and independent of each other with respect to their local data management and ownership. The Addenbrookes and Udine hospitals have locally managed databases of mammograms, with several thousand cases between them. As part of the MGVO, registered clinicians have access to (suitably anonymised) mammograms, results, diagnosis and imaging software from other centres. Access is coordinated by the MGVO central node at CERN. See Figure 1.

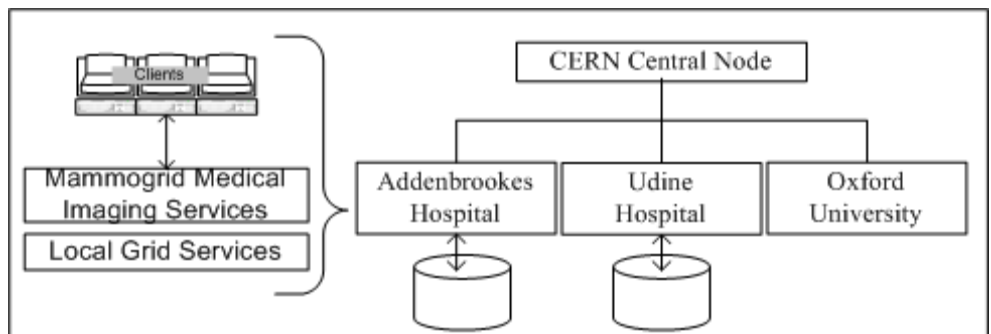

*Figure 1 – MammoGrid Virtual Organisation*

The adopted Grid implementation is the ALICE Environment (AliEn) [4] component of the EGEE-gLite middleware. gLite [5] is the middleware for grid computing of the EU-funded EGEE project [6]. The following AliEn services are used in the MammoGrid software:
a) *Authentication* for checking user credentials
b) *Resource broker* for resource management
c) *Storage Element* for file management
d) *Computing Element* for process management
e) *File transfer* for transferring files between nodes

A service-based approach to federate distributed mammography databases is employed [7]. A service-oriented approach permits the interconnection of communicating entities, called services, which provide some capabilities through exchange of messages. The services are 'orchestrated' in terms of service interactions – how services are discovered, how services are invoked, what can be invoked, the sequence of service invocations, and who can execute. The MammoGrid Services (MGS) are a set of services for managing mammography images and associated patient data on the grid.

The MGS are:
a) *Add* for uploading files (DICOM [8] images and structured reports) to the grid system
b) *Retrieve* for downloading files from the grid system
c) *Query* for querying the federated database of mammograms
d) *AddAlgorithm* for uploading executable code (e.g. SMF, CADe) to the grid system
e) *ExecuteAlgorithm* for executing grid-resident executable code on grid-resident files on the grid system
f) *Authenticate* for logging into the MGVO

Figure 2 illustrates the services that make up the MGVO. For simplicity, the Oxford University is not included.

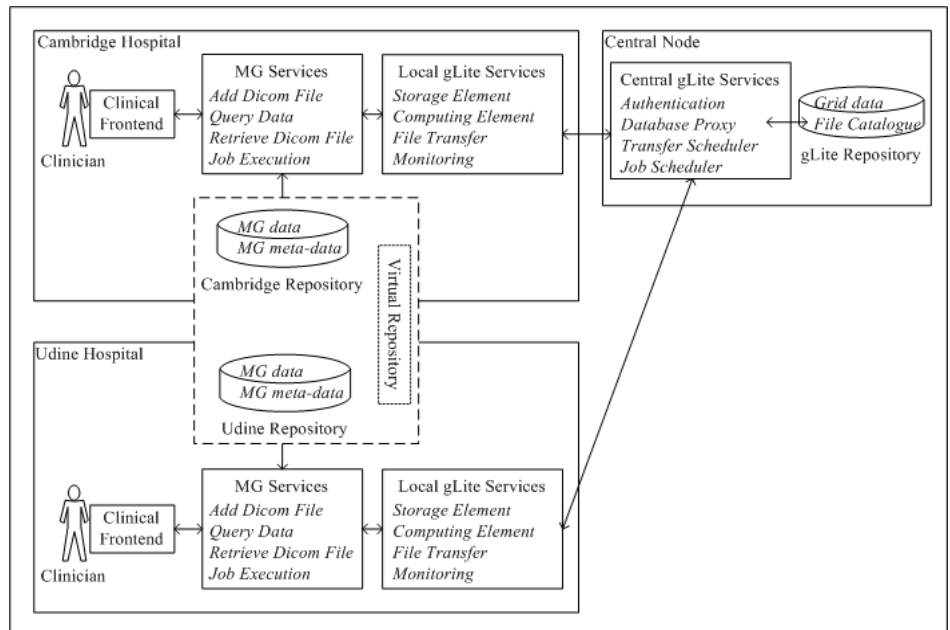

Figure 2 – MammoGrid Services

## 4. Query Handling

The MammoGrid project aims to provide a proof-of-concept demonstrator that allows clinicians to analyze mammograms resident on a Grid infrastructure. Clinicians define their mammogram analysis in terms of queries they wish to be resolved across the collection of data repositories. Queries can be categorized into simple queries (mainly against associated data stored in the database as simple attributes) and complex queries which require derived data to be interrogated or an algorithm to be executed on a (sub-)set of distributed images. The important aspect is that image and data distribution are transparent for radiologists so queries are formulated and executed as if these images were locally resident.

Queries are executed at the location where the relevant data resides, i.e. sub-queries are moved to the data, rather than large quantities of data being moved to the clinician, which can be prohibitively expensive given the quantities of data. Figure 3 illustrates how queries are handled in MammoGrid.

The *Query Analyzer* takes a formal query representation and decomposes into (a) a formal query for local processing, and (b) a formal query for remote processing. It then forwards these decomposed queries to the *Local Query Handler* and the appropriate *Remote Query Handler* for the resolution of the request. The *Local Query Handler* generates query language statements (e.g. SQL) in the query language of the associated Local DB (e.g. MySQL). The result set is converted to XML and routed to the Result Handler. The *Remote Query Handler* is a portal for propagating queries and results between sites. This handler forwards the formal query for remote processing to the Query Analyzer of the remote site. The remote query result set is converted to XML and routed to the Result Handler. See also [9].

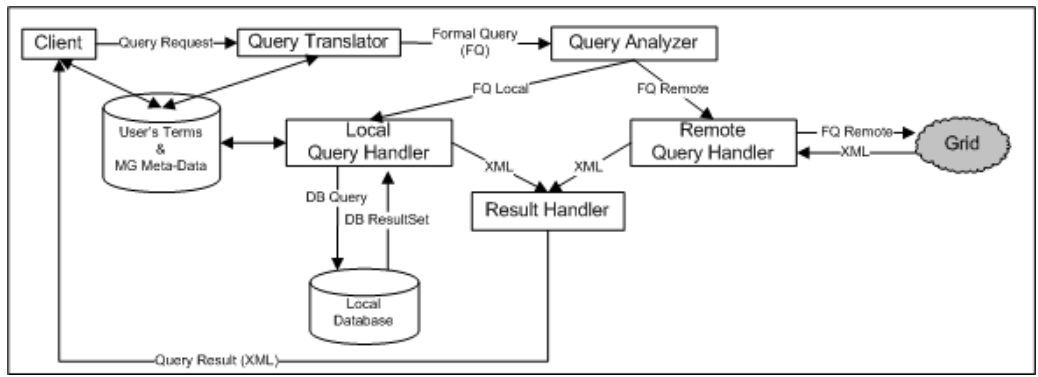

*Figure 3 – Query Handling*

## 5. Results

As of this writing the MGVO holds:

| Site | Number of Patients | Number of Image Files | Number of SMF Files | Associated Database Size | File Storage Size |
|---|---|---|---|---|---|
| Cambridge | 813 | 2798 | 2738 | 29 Mb | 70 Gb |
| Udine | 489 | 4663 | 2372 | 37 Mb | 85 Gb |

The average processing time for the core services are: (1) *Add* 8Mb DICOM file takes 7 seconds (2) *Retrieve* 8Mb DICOM file from a remote site takes 14 seconds (3) SMF workflow of *ExecuteAlgorithm* and *Add* takes 200 seconds. For querying:

| Query | Cambridge | Udine | Num images | Num patients |
|---|---|---|---|---|
| By Id: Cambridge patient | 2.654 sec | 2.563 sec | 8 | 1 |
| By Id: Udine patient | 2.844 | 3.225 | 16 | 1 |
| All female | 103 | 91 | 12571 | 1510 |
| Age [50,60] and ImageLaterality=L | 19.489 | 22.673 | 1764 | 357 |

Clinicians are still in the process of scanning and annotating cases contributing to several ongoing studies. These include (1) Cancers versus Control study: breast density study using SMF standard (2) Dose/Density study: exploring the relationship between mammographic density, age, breast size and radiation dose (3) CADe and validation of SMF in association with CADe. These studies aim to show how health professionals can work together without co-locating, and the MammoGrid approach provides a suitable environment for new forms of clinical collaboration.

## 6. Conclusions

The MammoGrid project has recently delivered its first proof-of-concept prototype enabling clinicians to store mammograms along with appropriately anonymised patient meta-data and to provide controlled access to mammograms both locally and remotely stored. A typical database comprising several thousand mammograms is being created for user tests of the query handler. The prototype comprises a high-quality clinician visualization workstation used for data acquisition and inspection, a DICOM-compliant interface to a set of medical services (annotation, security, image analysis, data storage and querying services) residing on a so-called 'Grid-box' and secure access to a network of other Grid-boxes connected through Grids middleware.

The MammoGrid Virtual Organisation is a distributed computing environment for harnessing the use of and access to massive amounts of mammography data across Europe. The MammoGrid approach uses grid technologies, service-orientation and database management techniques to federate distributed mammography databases allowing healthcare professional to collaborate without co-locating.

In the last couple of years, several Grid projects have been funded on health related issues at national and European levels (e.g. [10], [11]). These projects have a limited lifetime, from 3 to 5 years, and a crucial issue is to maximize their cross fertilization. Indeed, the HealthGrid [12] is a long term vision that needs to build on the contribution of all projects. The HealthGrid initiative, represented by the HealthGrid association, was initiated to bring the necessary long term continuity. Currently the HealthGrid Association is compiling an extensive White Paper on the requirements for Grids in biomedical healthcare [13].

Grid computing is a promising distributed computing paradigm that can facilitate the management of federated medical images. This technology spans locations, organizations, architectures and has the potential to provide computing power, collaboration and information access to everyone connected to the Grid. Grid-based applications like the MammoGrid project benefit from this solution being based on open-internet standards. These applications are potentially cross platform compatible, cross programming interoperable and widely accepted, deployed and adopted.

## 7. Acknowledgments

The authors thank their home institutes, the European Commission and acknowledge the support of the following MammoGrid Collaboration members: Prof R. Amendolia (formerly CERN), T. Solomonides, T. Hauer, D. Manset, W. Hassan (UWE), Dr R. Warren, I. Warsi (Addenbrookes Hospital), Dr C. Del Frate (Udine Hospital), Prof. M. Brady, C. Tromans (Oxford Univ.), M. Cordell, T. Reading (Mirada Solutions), Drs E. Fantacci (Pisa Univ.)& P. Oliva (Sassari Univ.), Dr A. Retico (Torino) and Dr P. Buncic and P. Saiz (CERN,AliEn).

**Address for correspondence:**

Dr Florida Estrella, DSU-TT Group, CERN, Geneva 23, Switzerland 1211

Florida.Estrella@cern.ch